\documentclass[useAMS,fleqn,usenatbib]{mn2e}
\usepackage{times}
\usepackage{amsmath}
\usepackage{bm}
\usepackage{url}
\usepackage{graphicx}
\usepackage{hyperref}
\input amssym.tex

\mathchardef\mhyphen="2D

\newcommand{\hmpc}{\,$h^{-1}$\,Mpc}

\newcommand{\chimp}{$(h^{-1}\, {\rm Mpc})^3$}

\newcommand{\camb}{{\scshape camb}}

\newcommand{\eg}{e.g.\ignorespaces }

\newcommand{\ik}{I_{\rm K}}
\newcommand{\Sk}{S_{\rm K}}
\newcommand{\rdet}{r_{\rm deterministic}}
\newcommand{\ikprim}{\ik^{\rm prim}}
\newcommand{\ikz}{\ik^{z=0}}
\newcommand{\rcut}{r_{\rm cut}}

\newcommand{\msun}{M_\odot}
\newcommand{\rhobarm}{\bar{\rho}_{\rm m}}

\chardef\til=`\~

\begin{document}

\title[Primordial complexity]{Kolmogorov complexity in the Milky Way and its reduction with warm dark matter}

\author[Neyrinck]
{Mark C.\ Neyrinck$^1$\thanks{E-mail:neyrinck@pha.jhu.edu}\\
$^1$Department of Physics and Astronomy, The Johns Hopkins University, Baltimore, MD 21218, USA\\
}

\maketitle

\begin{abstract}
We discuss the Kolmogorov complexity of primordial patches that collapse to form galaxies like the Milky Way; this complexity quantifies the amount of initial data available to form the structure. We also speculate on how the quantity changes with time.  Because of dark-matter and baryonic collapse processes, it likely decreases with time, i.e.\ information sinks dominate sources.  But sources of new random information do exist; e.g., a central black hole with an accretion disk and jets could in principle broadcast small-scale quantum fluctuations over a substantial portion of a galaxy.

A speculative example of how this concept might be useful is in differentiating between warm (WDM) and cold (CDM) dark matter. With WDM, the initial patch that formed the Milky Way would have had few features, making the present high degree of structure a curiosity.  The primordial patch would have had only several billion independent information-carrying `pixels' if the WDM particle had a mass of 1 keV. This number of `pixels' is much less than even the number of stars in the Milky Way.  If the dark matter is proven to be warm, the high degree of structure in the Milky Way could have arisen in two ways: (1) from a high sensitivity to initial conditions, like an intricate fractal arising from a relatively simple computer code; or (2) from random information generated after the galaxy formed, i.e.\ not entirely deterministically from the initial conditions.
\end{abstract}

\begin {keywords}
  large-scale structure of Universe -- cosmology: theory
\end {keywords}
\section{Introduction}
Intuitively, any sensible measure of the complexity, or information content, of the Milky Way, would be staggeringly vast, compared to that usually encountered by humans.  A convenient definition to use is $\ik$, the Kolmogorov complexity \citep{Kolmogorov1963}, which is the length (e.g.\ in digital bits) of the shortest-possible algorithm that describes the object. This description can be considered to consist of two parts: a program, and data fed to the program.

How to estimate $\ik$ for the Galaxy? A full description would include all positions and momenta of all particles, to the accuracy allowed by the uncertainty principle. This description would carry a huge amount of raw data, with no processing necessary. But this description can surely be compressed.  Particles form organized structures, like molecules, crystals, lifeforms, planets, stars, star clusters, and spiral arms. But we aim at an exact description; explicitly encoding descriptions of such structures with all their possible variations rapidly becomes unsatisfying and overwhelming. 

In this paper, we do not specify a method for obtaining a minimal description giving $\ik$ from an evolved density field, but as Kolmogorov showed, one does exist. If the formation of structure in the Galaxy is deterministic, perhaps a minimal description would consist of the initial conditions, together with an algorithm, i.e.\ the set of physical laws that formed the Galaxy from these initial conditions. In the current cosmological paradigm, the primordial patch that formed the Milky Way was imprinted with a random pattern of scalar density fluctuations during inflation, in the first instants after the Big Bang. We call this patch, and the tidal field, the `initial conditions.' Further degrees of freedom were present initially in the form of tensor modes (present on large scales) and decaying modes, but these are likely negligible for galaxy formation. 

On large, linear scales, the dynamics seem to be deterministic, but at what scale does a one-to-one correspondence between initial and final conditions fail? When it fails, does a volume of initial-conditions phase space generally shrink or expand in the final-conditions phase space? That is, do information sinks or sources dominate? A common attitude is that if sufficient resolution and computational resources existed for a simulation, there would be a one-to-one correspondence between initial and final conditions. Since we are still far from able to resolve all relevant processes, the question of at what scale structure in the Galaxy is deterministic is not yet a crucial one for modeling, but the question is still of interest theoretically and philosophically.

\subsection{Information sinks}
Information `sinks' occur when many sets of initial conditions (microstates) correspond to indistinguishable final conditions (macrostates). This can be used to define an information entropy, the logarithm of the number of microstates per macrostate. Neglecting sources, the information $\ik$ decreases with time. This deficit $\Sk=\ikprim-\ik$ between the initial and final information is a sort of `entropy.'  A perfect information sink is a black hole, which carries physical entropy. It is possible that $\Sk$ relates somehow to physical entropy, but here we claim no correspondence. 

On extragalactic scales, non-linear, collisionless dark-matter dynamics give an information sink.  This can be seen in the reduction with time in power-spectrum Fisher information on small scales \citep[e.g.][]{RimesHamilton2006}. This Fisher information gives essentially the number of statistically independent Fourier modes in a field. Much of this can be recovered by a change of variables to the log-density \citep{NeyrinckEtal2009}, but the restoration is not complete: fluctuations that were once imprinted on a structure that has collapsed are still completely lost, when resolved on a cosmological scale. See \citet{NeyrinckYang2013} for a discussion of how information imprinted at different scales behaves globally, in different density variables.

However, a Fourier-space description of a non-periodic patch is awkward. More intuitively, consider a patch of real-space pixels, of sufficient resolution to capture all physical fluctuations present in the initial density field. If some initial pixels become irrelevant in determining the final structure smoothed over some scale, then they may be omitted (or encoded with fewer bits than others) in a minimal description of the patch; this would reduce $\ik$. It is unclear whether primordial information is truly destroyed in this manner by non-linear dynamics on subgalactic scales, but the Fourier-space analysis described in the previous paragraph suggests that it is.

Baryonic physics likely give further sinks, beyond that of collisionless dark-matter dynamics. In the linear regime, baryonic fluctuations are the same as the dark matter, except further damped by photon diffusion and gas pressure forces \citep[\eg][]{Silk1968,GnedinVanVliet2003}. So, in the linear regime, there is not additional information present in the baryons that is not in the cold dark matter. Even for slightly nonlinear objects like filaments, the gas is a smoothed version of the dark matter \citep{HarfordHamilton2011}. Eventually, within galaxies, dissipation and cooling in the baryons allows much more clustering than in the dark matter \citep{RuddEtal2008}, but, by definition, until non-primordial randomness (see the next section) acts, that structure is still entirely determined by the initial dark-matter fluctuations. Indeed, dissipation causes $\ik$ to decrease when e.g.\ a star forms, since the end result has little dependence on fluctuations well inside the patch that formed the star. 

\subsection{Information sources}
Information `sources' are processes that inject new, non-primordial randomness into the system. Each time an atom radiates a photon from an excited energy state, it is a probabilistic process, and the randomness in the time and direction of photon emission adds information to the system. Because these quantum processes occur far into the large-number regime for most astrophysical processes, information sources are negligible after coarse-graining on an astronomical scale.

However, there are many instabilities in astrophysical processes that can amplify small fluctuations chaotically. There are also processes that propagate structure on very small scales to large scales, e.g.\ supernovae and jets.  The fine-scale structure of a supernova explosion, perhaps partially arising from microscopic quantum randomness, can grow to scales comparable to a galaxy, especially for small galaxies. Active galactic nuclei (AGN), and other jets, are another possible non-primordial information source, potentially broadcasting small fluctuations within a few Schwarzschild radii of a black hole to much larger scales.  AGN are known to affect the power spectrum on non-linear scales \citep{LevineGnedin2006}; they can evidently affect density profiles in galaxies \citep{PontzenGovernato2012} and even clusters \citep{vanDaalenEtal2011}. However, it is not clear on what scale non-primordial randomness affects the structures these processes produce; it is plausible that non-primordial randomness only affects fine details, and not, for example, density profiles well away from the AGN.  But it could be that non-primordial randomness accumulates over time, eventually dominating the information budget on all galactic, or cluster, scales.  The structure on Earth, for example, may be entirely determined by non-primordial randomness.

\subsection{Information conservation and warm dark matter} 
In this paper, we explore the consequences of conservation (or, if anything, net destruction) of $\ik$ from initial to final conditions. This could plausibly hold on scales larger than a critical scale, $\rdet$. This scale could be position-dependent; in a void, far away from processes broadcasting non-primordial randomness, $\rdet$ could be very small, but near a galaxy, it may be of order the galaxy's radius.

Taking the Milky Way as an example, we leave $\rdet$ unspecified, but assume for argument that it is less than the scale of the Galaxy.  After coarse-graining at $\rdet$, we assume that $\ikz \le \ikprim$, where $\ikz$ is the Kolmogorov complexity of the Galaxy at $z=0$, and $\ikprim$ is the Kolmogorov complexity of the initial conditions on which the present structure depends. We assume that $\rdet$ is large enough to smooth over thermal motions, but do not explicitly use a particular scale below, since it is so uncertain. $\ik$ should include the information necessary to encode the algorithm used to convert raw data into the present structure, $\ik^{\rm physics}$, but we neglect this compared to $\ikz$ and $\ikprim$. Note that often in computer science, often the dominant part of $\ik$ is in the algorithm, not the data, as it is here.  If $\ikprim$ is surprisingly small compared to a naive estimate of the final complexity $\ikz$, that is like an apparently complex fractal with hidden simplicity, indicating a high sensitivity to initial conditions. Some of this sensitivity may be quite subtle, e.g.\ the reionization provided by the first stars, which form at locations determined by the initial conditions. Gas may become ionized cosmologically faraway from these first stars, inhibiting small galaxies from forming.  

In \S \ref{sec:ikprim}, we show how the primordial information depends on small-scale power cutoffs. In particular, if the dark matter was warm, i.e.\ slightly relativistic at decoupling, then primordial fluctuations were damped. In \S \ref{sec:discussion}, we discuss how this primordial information tally might be used, and conclude.

Of course, a lack of small-scale power from warm dark matter (WDM) would show up in other ways, more directly observable than information content, but all related to the smoothing of initial power that underlies our information-content investigation. WDM reduces the population of Milky Way satellites compared to CDM; it was the `missing satellites' problem with CDM that first prompted the recent burst of interest in WDM \citep[\eg][B01]{BodeEtal2001}. However, currently allowed WDM masses do not seem to solve these issues by themselves \citep{SchneiderEtal2014}. \citet{WeinbergEtal2013} give a recent review of the status of these issues. \citet{HoganDalcanton2000} also investigate how WDM and dark-matter self-interaction influence the structure and stability of dark-matter haloes. There have also been several proposals to look at structure within the Milky Way, its stellar streams, and its arrangement of satellites for similar cosmological information \citep[\eg][]{LyndenBells1995,MetzEtal2009,StarkenburgEtal2009,LiHelmi2009,LawMajewski2010,CooperEtal2011,BozekEtal2013,NganCarlberg2014}. \citet{FreemanBlandHawthorn2002} give an overview of observational knowledge about the primordial patch that formed the Milky Way, and its structure and accretion history.

\section{Quantifying the primordial information of the Milky Way}
\label{sec:ikprim}
Observations of the cosmic microwave background (CMB) constrain the primordial Universe to have very nearly Gaussian, small density fluctuations \citep[\eg][]{PlanckNonGaussianity}. The initial pattern of fluctuations, if pixelized, can be considered to consist of independent random Gaussian densities in each pixel, with the resulting field multiplied in Fourier space by $\sqrt{P_{\rm init}(k)}$, the square root of the initial power spectrum. The statistical independence of each pixel implies that the primordial information is incompressible, at least if power is substantial at all scales. Then the initial information at resolution $r$ of a primordial patch $\ikprim(r)=bN_{\rm pix}(r)$, where $N_{\rm pix}(r)$ is the number of pixels of size $r$ necessary to cover the patch, and $b$ is the (arbitrary) number of bits used to encode each pixel. In comparing final to initial complexity, we might alleviate $b$'s arbitrariness by considering the precision with which initial conditions need to be specified to obtain a given precision in the final conditions, in a similar spirit as a Lyapunov exponent. But here, we simply take $b$ to be a constant, necessary to introduce because we treat continuous system digitally. See the Appendix for a bit more discussion of this issue.

Note that enumerating information in volume pixels is at odds with the holographic principle: at high energies, the amount of information in a volume is proportional to its surface area, not volume \citep[for a review, see][]{Bousso2002}. This distinction is crucial for black holes. At high energies, if Planck volumes are considered to be the fundamental information-carrying units, counting the number of them in a small patch gives an information overestimate, because many available high-energy arrangements of states would give black-hole collapse \citep{Bousso2002}. Here, however, we consider lower energies; a fiducial epoch to measure $\ikprim$ is when the CMB was emitted. The $10^{-5}$ fluctuations in the modest-density plasma present there are non-relativistic. We also consider galaxy-forming patches much smaller than the cosmological horizon. So, we adopt a count by volume instead of surface area.

Power should not exist at arbitrarily small scales primordially. By `primordial' here we mean at an epoch after inflation, and after free-streaming thermal motions had finished smoothing out dark-matter fluctuations \citep[\eg][]{BondSzalay1983}, but while fluctuations in the dark matter density field were still small enough to be entirely in the linear regime. This smoothing process imparts a cutoff to the power spectrum at a comoving (all scales mentioned here are comoving) length scale depending on the dark-matter mass. If power is zero on scales less than $\rcut$ ($k\gtrsim2\pi/\rcut$ in Fourier space), $\ikprim$ may be easily estimated for a virialized structure of mass $M$. The number of pixels is just the initial volume of the patch ($M/\rhobarm$), divided by the pixel volume, so
\begin{equation}
 \ikprim = \frac{Mb}{\rhobarm\rcut^3}=\frac{M}{1.1\times 10^{11} \msun} \frac{\Omega_{\rm DM}}{0.27}  \left(\frac{1\, h\, {\rm Mpc}}{\rcut}\right)^3b,
 \label{eqn:ikprim}
\end{equation}
where $\rhobarm$ is the mean matter density (i.e.\ the conversion factor between mass and Lagrangian volume occupied in the initial conditions), $\Omega_{\rm DM}$ is the fraction of the critical density comprised of dark matter, and $b$ is the number of bits required to encode each pixel. Note that $\ikprim$ decreases steeply with $\rcut$, implying high sensitivity to this parameter if it could be measured. Birkhoff's theorem implies that distant fluctuations can only affect local motions in a roughly spherical collapsing patch through the tidal field \citep{PeeblesBook1993,DaiEtal2015}. Larger-scale modes can also produce a large-scale overdensity or underdensity, which would affect the rate of structure formation in the patch. But it is a reasonable assumption that the information determining its final structure was entirely contained in this Lagrangian patch, plus a negligible set of extra quantities to specify the large-scale density and tidal field.

Complicating the picture is the exact form of a dark-matter streaming cutoff in power, which would not be completely sharp in either Fourier or configuration space. For a neutrino-like WDM particle, B01 obtain a polynomial fit over scales near $\alpha$,
\begin{equation}
P_{\rm WDM}(k) = P_{\rm CDM}(k)[1+(\alpha k)^{2}]^{-10},
\label{eqn:cutform}
\end{equation}
where $\alpha$, a characteristic cutoff scale, is given by
\begin{equation}
\alpha\approx0.05\left(\frac{\Omega_{\rm DM}}{0.4}\right)^{0.15}\left(\frac{h}{0.65}\right)^{1.3}\left(\frac{\rm keV}{m_{DM}}\right)^{1.15} h^{-1}{\rm Mpc}.
\label{eqn:alpha}
\end{equation}

\begin{figure}
  \begin{center}
    \includegraphics[width=\columnwidth]{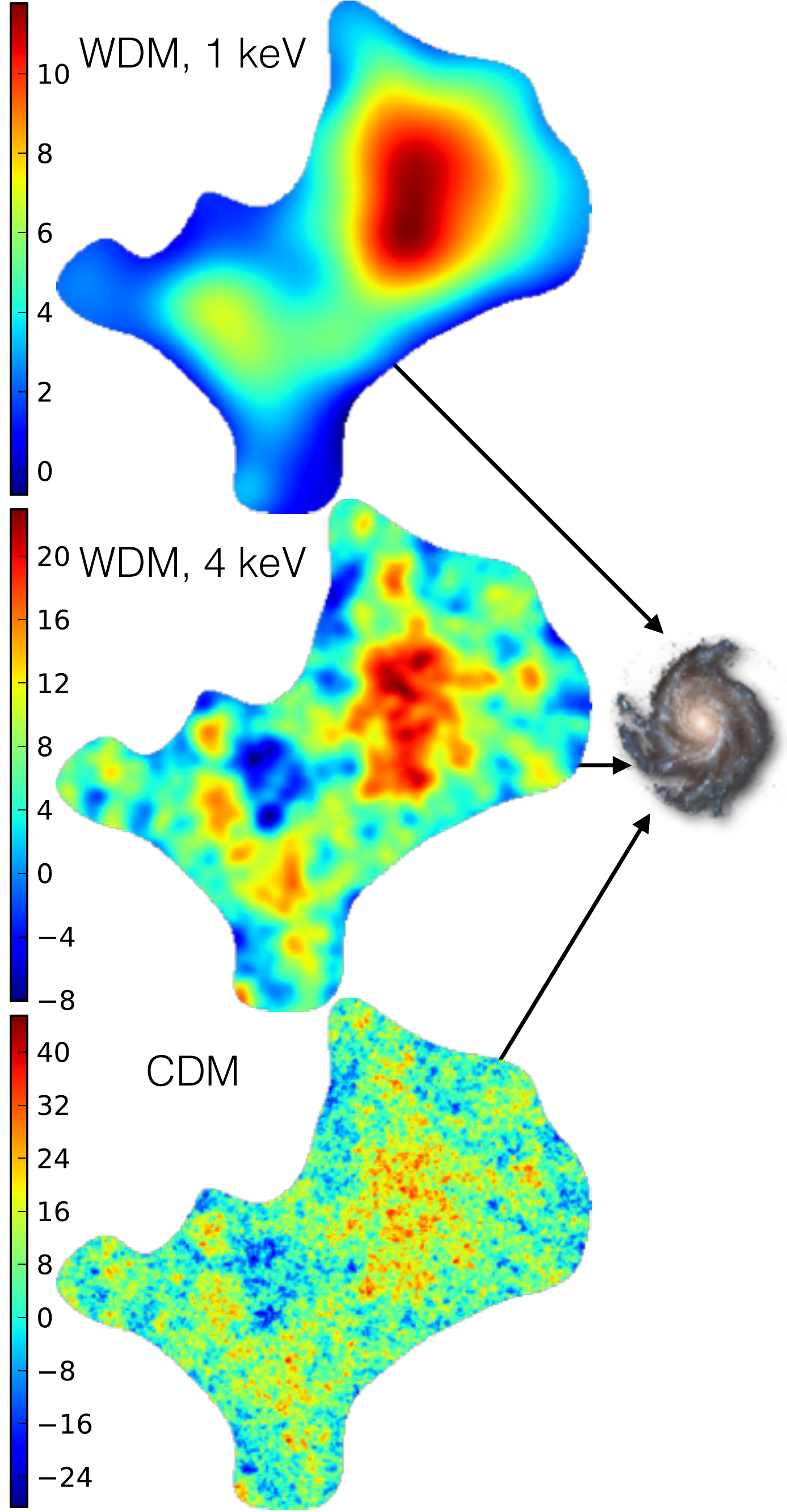}
  \end{center}  
  \caption{Initial-conditions (Lagrangian) patches that might collapse to form the Milky Way, assuming three different dark matter possibilities. The patch, color-coded by overdensity (linearly extrapolated to $z=0$), is a 2D slice of a high-densitiy 3D blob with about the mass of the Milky Way (see text for details). Pixels have side length 0.01 comoving \hmpc, the value of $\alpha$ in a 4 keV model (middle panel). Note that even at 4 keV, there is noticeable smoothing on larger scales than $\alpha$. There is very little structure visible in the 1 keV blob. In the CDM case, the structure in the blob continues to much smaller scales than the pixels.
  }
  \label{fig:mwcontract}
\end{figure}

For CDM, the dark-matter power cutoff is much smaller than Galactic scales, $\sim0.6$ pc for a 100 GeV neutralino, or even $\sim 0.003$ pc for an axion \citep{DiemandEtal2005}. For WDM, however, $\rcut$ may be comparable to Galactic scales. In a fiducial WDM case suggested to ameliorate small-scale problems in CDM (B01), $\alpha=0.05$\hmpc, corresponding to a WDM particle mass of 1 keV. However, recent analyses of Lyman-$\alpha$ fluctuations and ultra-faint dwarf galaxies to simulations favor a larger dark-matter mass, $m_{\rm DM} \gtrsim 4$ keV \citep{PolRic2011,VielEtal2013}. A 4-keV WDM particle gives $\alpha=0.01$\hmpc, using $\Omega_{\rm DM}=0.27$, $h=0.7$, in Eq.\ (\ref{eqn:alpha}).

To make progress in quantifying $\ik$, we assume that the cutoff is steep enough to erase structure to zero at the effective sharp cutoff scale $\rcut$, which is only an order of magnitude or two smaller than $\alpha$. Specifying a fiducial precision of $10^{-6}$ that corresponds to single precision gives $\rcut\approx\alpha/50$ for cutoffs of the form in Eq.\ (\ref{eqn:alpha}). See the Appendix for details.

Including dark matter, the Milky Way halo has mass $M_{\rm MW}\approx10^{12} \msun$ \citep[\eg][]{McMillan2011}, giving a Lagrangian volume of $\sim 8$ \chimp. For a 1-keV particle, setting $\rcut=10^{-3}$\hmpc, $\ikprim=8$ gigapixels, fitting into a cube 2,000 pixels on a side (a pixel's density is encoded with $b$ bits). For familiarity, we assume single-precision, 32-bit floating-point numbers, although this is arbitrary. Then these pixels could be encoded in $\sim 30$ gigabytes. $N_{\rm pixels}\propto \alpha^{-3}\propto m_{\rm DM}^{3.45}$. For a 4-keV particle, $\rcut=2\times10^{-4}$\hmpc, and $\ikprim=1,000$ gigapixels, fitting into a cube 10,000 pixels on a side. This could be encoded at single precision with 4 terabytes. In contrast, in a CDM scenario, with a characteristic cut at $\rcut\approx0.012$ pc (assuming the same factor of 50 as above, although the cutoff may have a different shape), $\ikprim$ enlarges to $\sim10^{24}$ pixels, encodable at single-precision in $\sim 10^{25}$ bytes, or $10$ yottabytes.

Fig.\ \ref{fig:mwcontract} shows patches of Gaussian random fields, each with the same Fourier phases, that might collapse to form the Milky Way, in different dark-matter scenarios. According to the above estimate, the smoothest, 1 keV blob is specified with 2000$^2$ pixels. Visually, there is so little structure that the field seems specifiable with many fewer pixels than that, suggesting that this may be an overestimate of the true information content. Quantifying $\ikprim$ unambiguously is necessary to make precise constraints.

These blobs are slices through a 512$^3$ Gaussian random field with a power spectrum generated with \camb\ \citep{camb} using the parameters given above. The blob shown exceeds an average spherical-collapse overdensity $\delta=1.69$. The contours outlining the blob are isodensity contours in the field after applying a 1\hmpc-dispersion Gaussian filter to the CDM field, with the level of the contour set to give a Milky-Way-mass object.

\section{Discussion}
\label{sec:discussion}
We argued that if the dark matter is a 1-keV WDM particle, then the primordial patch that formed the Milky Way contained on the order of 8 gigapixels of information; if the dark matter is of mass 4 keV, this increases to $\sim 1000$ gigapixels. If the Milky Way can be shown to have more information $\ikz$ in its structure than these, in a way that excludes random information introduced after the primordial fluctuations were imprinted, this could in principle constrain a power cutoff $\rcut$, sensitive for example to dark-matter warmth. This method may not be as rigorous or discriminating as other tests in constraining the WDM, but it would still be curious if the high degree of structure in the Galaxy was able to come from a small amount of primordial information, as in a WDM scenario.

What are the prospects for estimating $\ikz$ from the Galaxy's current structure? While much of this paper has discussed $\ik$ in terms of primordial information that is evolved as in a simulation, $\ik$ may alternatively be accessible by analyzing the current structure, especially if $\ik$ decreases with time. A naive estimate would assign some information to each structure that might have some dependence on the initial conditions, such as a star. There are $10^{11-12}$ stars in the Galaxy \citep[\eg][]{FranckEtal2001}, depending on what one counts as a `star.' This number far exceeds the $\sim 10^9$ primordial pixel count for a 1-keV WDM particle, and is of the order of the $\sim 10^{12}$ pixel count at 4 keV.  Under the strong assumption that each star contains one independent unit of information of comparable information content to a primordial pixel of size $\rcut$ (i.e. that $b$, the `number of bits per pixel' above, is the same in both cases), and that Galactic structure formation is deterministic from the initial conditions on scales of typical stellar separation, this suggests that the high degree of apparent complexity in the Milky Way makes low-mass WDM less plausible. Here, we assume $\rdet\sim 1$ pc, some typical interstellar scale.

Of course, the assumption that each star has a unit of independent information is poor; in reality, many stars originate in clusters and giant molecular clouds, causing much of their information to be degenerate. Also, much information about the assembly of the Galaxy lies in the character of its stellar populations, metallicities, etc. Primordial information may be encoded in these patterns. There is also independent information in the arrangement of gas and dark matter, that we have excluded.

What could make the comparison of initial and final information more meaningful and rigourous? An uncertainty is the degree to which information is lost on subgalactic scales through nonlinear gravitational and hydrodynamic processes. In principle, a way to test this is to vary the initial conditions slightly (e.g. by amplifying a set of Fourier modes, or changing the initial density in a set of pixels), and observe the result. One would have to be entirely sure that the simulation would not introduce random artifacts, however; thus, this test may only be practical on relatively large subgalactic scales. Also, state-of-the-art WDM hydrodynamic simulations would help to identify where the underlying primordial simplicity in WDM might show up. WDM presents special difficulties in simulations \citep{WangWhite2007,AnguloEtal2013}. Still, the morphology of star formation is different in a WDM scenario \citep{GaoTheuns2007,GaoEtal2015}: stars form in cosmic-web filaments rather than in conventional galaxies.  In these simulations, a Milky Way protogalaxy shows little structure in WDM, but plentiful structure in CDM. WDM should also impart simplicity in the way the stars and gas occupy position-velocity six-dimensional phase space.  The `catastrophes' \citep{ArnoldEtal1982,HiddingEtal2014} which initially formed the Galaxy, or protogalaxies, may still be present, perhaps detectable in {\it Gaia} data.

If the dark matter is proven to be warm, e.g.\ through direct detection, the rich structure in the Galaxy could have only two explanations, both surprising philosophically (to the author).

First, mechanisms in star formation and hydrodynamics could allow fine-scale structure to develop even from smooth initial conditions, in a deterministic way. This would be like an apparently highly complex fractal, generated from a simple algorithm. This would imply that the number of possible `Milky Ways' is $\sim 2^{\ik}$. This is large, but still curiously finite.

The second possibility is that Galactic structure is dominated by non-primordial randomness, e.g.\ supplied in microscopic processes operating during star formation, or in AGN, which can broadcast small-scale fluctuations to large scales. If the first possibility can be ruled out with WDM simulations including full hydrodynamics, this would suggest that the Milky Way's structure is not deterministic purely from the initial conditions.

\section*{Acknowledgments}
I thank Miguel Arag\'on-Calvo, Brandon Bozek, Istv\'an Szapudi, Xin Wang, Donghui Jeong, and Alex Szalay for helpful discussions. I am grateful for support from a New Frontiers in Astronomy and Cosmology grant from the Sir John Templeton Foundation, and a grant in Data-Intensive Science from the Gordon and Betty Moore and Alfred P. Sloan Foundations.

\bibliographystyle{mn2e}
\bibliography{refs}

\section*{Appendix: smooth power cutoffs}
One way to specify $\rcut$ from a parameter like $\alpha$ is with a precision to which the initial density field should be known. $\rcut(b)$ can be defined as the critical scale where additional pixel fluctuations from scales smaller than $\rcut$ are unresolved at the precision specified by $b$, the number of bits used to represent each pixel density.  The size of these fluctuations can be quantified by $\sigma^2(r)$, the variance in cells of radius $r$; this may be computed from an integral over $P(k)$ multiplied by the square of the spherical top-hat (using spherical cells for simplicity) pixel window function. 

The fiducial fractional precision we use for a pixel density, to determine the effective sharp $\rcut$, is ${\rm Err}(\rcut)\equiv[\sigma(0)-\sigma(\rcut)]/\sigma(0)=10^{-6}$. Attenuating a \camb\ linear power spectrum (generated using the above concordance cosmological parameters) with the cutoff in Eq.\ (\ref{eqn:cutform}), we found that ${\rm Err}(\rcut)=10^{-6}$ at $\rcut\approx\alpha/50$, for $\alpha\lesssim1$\hmpc. 

This choice of $10^{-6}$ is admittedly quite arbitrary, but there are some reasons for the choice. It is the precision at which representational discreteness becomes visually obvious in ${\rm Err}(r)$, log-log plotting it as a function of $r$, if $\sigma(r)$ is encoded with 32-bit floating-point single precision. $10^{-6}\approx2^{-20}$ is 16 times the fundamental precision of the mantissa at single precision, which is encoded with 24 bits. Going all the way down to $2^{-24}$ could seem a more obvious choice at single precision, but we wish to allow resolved modes to able to contribute more than in the last couple of bits of precision. Another reason we do not assume higher precision is that we do not want the results to depend on precise knowledge of the cutoff shape for many orders of magnitude smaller than $\alpha$.

An undesirable aspect of the definition in Eq.\ (\ref{eqn:ikprim}) is the high sensitivity of $\ik$ to the sharp $\rcut$, which is strange for a smooth cutoff. One strategy to reduce this effect might be to encode a field of pixels in Fourier space, encoding each Fourier mode with a number of bits that depends on the mode's contribution to the total variance of the field ($\propto P(k)k^3$). Modes with higher $P(k)$ could be encoded with more bits.  This would give a gradual plateau in information as resolution is increased. However, leave this investigation to future work.

\end{document}